\documentclass[aps,prd,twocolumn]{revtex4-2}

\usepackage{amssymb}
\usepackage{amsmath}

\usepackage{graphics}
\usepackage[export]{adjustbox}

\usepackage{xparse}

\usepackage{hyperref}
\usepackage{hypernat}

\usepackage{layouts}

\usepackage[caption=false,labelfont=normalsize,labelformat=empty, textfont=normalsize,justification=centering]{subfig}

\renewcommand{\Re}{\mathrm{Re}}

\newcommand{\iu}{\mathrm{i}\hskip0.07em}

\begin{document}

\title{Modifying the Dyson series for unstable particles and resonances}

\author{Peter Mat\'ak}
\email{peter.matak@fmph.uniba.sk}
\affiliation{Department of Theoretical Physics, Comenius University,\\ Mlynsk\'a dolina, 84248 Bratislava, Slovak Republic}

\date{\today}

\begin{abstract}
We introduce a modification of the Dyson series based on perturbative unitarity as a starting point. The presented approach systematically avoids singularities and double-counting related to the presence of unstable particles as intermediate states and, at the same time, it does not rely on a specific choice of the renormalization scheme.
\end{abstract}

\maketitle

\section{Introduction}

In particle physics, either at colliders or in astroparticle calculations, some of the final states may be produced from decays of heavy unstable particles. At higher perturbative orders, new production channels occur, where the same unstable particles appear in diagrams as internal lines. Such contributions lead to singularities, which are cured when including finite widths originating from the imaginary part of the self-energy in a resummed propagator. The cure is not perfect -- the resonant enhancement reproduces the leading-order result, which seems to be accounted twice.  This double-counting is well known in the literature and often targeted with more or less physically motivated subtraction schemes, including explicit diagram removal \cite{Frixione:2008yi, Hollik:2012rc}, the exclusion of the near-resonance parts of the phase space \cite{Belyaev:1998dn}, or the diagram subtraction also known as the \textsf{Prospino} scheme \cite{Beenakker:1996ch, Plehn:2010gp, Gavin:2013kga}, in which a dedicated counter-term is introduced to remove the singularity. 

An accurate description of unstable particles requires more sophisticated approaches, especially for near-threshold calculations. Within the \emph{complex mass scheme}, the complex-valued counter-terms are used to remove the double-counting \cite{Denner:2005fg, Denner:1999gp, Denner:2006ic, Denner:2014zga}.
\emph{Unstable-particle effective theory}, on the other hand, relies on the systematic expansion in powers of small width-to-mass ratio \cite{Beneke:2003xh, Beneke:2004km, Beneke:2015vfa}. Both these approaches preserve gauge invariance.

This work introduces an alternative unitarity-based prescription leading to nonsingular results free of double-counting. Instead of modifying the perturbation theory by the expansion of finite-width propagators \cite{Tkachov:1997aq, Tkachov:1998uy, Nekrasov:2007ta, Nekrasov:2009nc}, we apply the cutting rules to forward diagrams with bare propagators only \footnote{By forward diagrams, we mean the Feynman diagrams with the same initial and final states.}. Propagators of unstable particles are cut following the standard Cutkosky rules as well. It is further shown that in comparison to the subtraction schemes mentioned above \cite{Frixione:2008yi, Hollik:2012rc, Belyaev:1998dn, Beenakker:1996ch, Plehn:2010gp, Gavin:2013kga}, additional finite terms are restored in the NLO corrections. From this perspective, it becomes clear that the source of the problems is how the Dyson series is used in the unstable-particle case. We suggest its modification to describe resonant particle production. However, in the beginning, the Dyson series is avoided altogether.

Employing a simple scalar model, it is demonstrated that neither singularity nor double-counting is present in a fixed-order calculation. The loop corrections to particle masses and LSZ reduction factors are obtained from cuts of forward diagrams. Those are carefully treated as distributions, as it is in modified perturbation theory \cite{Tkachov:1997aq, Tkachov:1998uy, Nekrasov:2007ta, Nekrasov:2009nc}. However, by cutting the forward diagrams, we always obtain polynomials in the coupling constant only, and no logarithms in couplings may appear in higher-order corrections \footnote{In modified perturbation theory, this may not be the case, as can be seen in Eq. (9) in Ref. \cite{Tkachov:1998uy}}.

\section{Unstable particles in fixed-order calculations}

Let us, for the sake of simplicity, consider a massive scalar field $\phi$ coupled to two massless scalars, $\varphi$ and $\eta$, through the Lagrangian density
\begin{align}\label{eq1}
\mathcal{L} \supset -\frac{1}{2}m\phi^2- \lambda_\eta\phi\eta\eta^* - \frac{1}{3!}\lambda_\varphi\phi \varphi^3.
\end{align}
We consider $\phi$ and $\varphi$ as real, while $\eta$ is complex. For a while, we assume that $\phi\rightarrow\eta\bar{\eta}$ is the only decay channel of $\phi$. We will comment on general branching ratios later. 

Assuming the collisions of two $\varphi$ particles, what is, at the leading $\mathcal{O}(\lambda^2_\varphi)$ order, the production rate of a single $\eta\bar{\eta}$ pair? Above the on-shell production threshold, the leading contribution will come from $\varphi\varphi\rightarrow \phi\varphi$ followed by the $\phi\rightarrow\eta\bar{\eta}$ decay. Therefore, the leading-order amplitude squared reads
\begin{align}\label{eq2}
\vert\mathcal{M}\vert^2_{\mathrm{LO}}(m^2) = \frac{\lambda^2_\varphi}{4\pi}\frac{s-m^2}{2s}
\end{align}
where the final-state phase space integration has already been performed. 

\begin{figure}
\centering\includegraphics[scale=1]{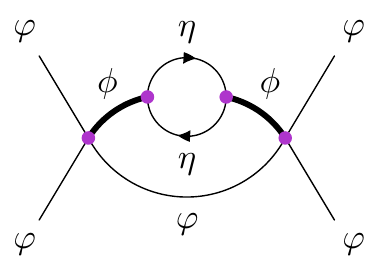}
\caption{\label{fig1} Forward scattering diagram generating the NLO corrections to $\eta\bar{\eta}$ production in $\varphi\varphi$ collisions according to Lagrangian density in Eq. \eqref{eq1}.}
\end{figure}

While working with \emph{zero-width propagators} only, the higher $\mathcal{O}(\lambda^2_\varphi \lambda^2_\eta)$ corrections to $\eta\bar{\eta}$ production should be obtained from cutting the forward diagram in Fig. \ref{fig1}. Let us start with the sum of the following two cut diagrams
\begin{align}\label{eq3}
\includegraphics[scale=1,valign=c]{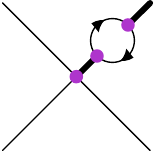}
\includegraphics[scale=1,valign=c]{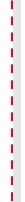}
\includegraphics[scale=1,valign=c]{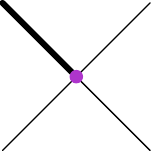}\hskip1mm+\hskip1mm
\includegraphics[scale=1,valign=c]{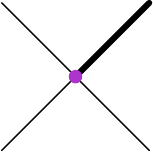}
\includegraphics[scale=1,valign=c]{cut.pdf}
\includegraphics[scale=1,valign=c]{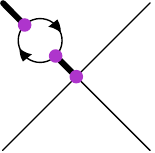}
\end{align}
where we suppress the particle labels. Here the diagrams to the right side of the cut come with complex conjugation. Denoting
\begin{align}\label{eq4}
-\iu\Sigma(k^2) = \includegraphics[scale=1,valign=c]{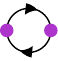}
\end{align}
the diagrams in Eq. \eqref{eq3} lead to
\begin{align}\label{eq5}
\int\limits^s_0 dk^2 \vert\mathcal{M}\vert^2_{\mathrm{LO}}(k^2) \delta(k^2-m^2)\,
2\Re\bigg[\frac{\Sigma(k^2)}{k^2-m^2+\iu\epsilon}\bigg].
\end{align}
Splitting the self-energy $\Sigma(k^2)$ into its real and imaginary parts, $\Sigma_{\mathcal{R}}(k^2)$ and $\Sigma_{\mathcal{I}}(k^2)$, respectively, and employing the identities
\begin{align}\label{eq6}
\frac{1}{k^2-m^2+\iu\epsilon} = \mathcal{P}\frac{1}{k^2-m^2}-\iu\pi\delta(k^2-m^2)
\end{align}
and \footnote{Similar approach has been used to study infrared finiteness by cutting box diagrams in Refs. \cite{Frye:2018xjj, Racker:2018tzw, Blazek:2021olf, Blazek:2021gmw}.}
\begin{align}\label{eq7}
2\delta(k^2-m^2)\mathcal{P}\frac{1}{k^2-m^2} = -\frac{\partial}{\partial k^2}\delta(k^2-m^2)
\end{align}
one can easily verify that the real part $\Sigma_{\mathcal{R}}(k^2)$ contributes to the NLO corrections by
\begin{widetext}
\begin{align}\label{eq8}
\vert\mathcal{M}\vert^{2,\phi}_{\mathrm{NLO}} = 
\bigg(\Sigma_{\mathcal{R}}(k^2)\frac{\partial}{\partial k^2} + \frac{\partial\Sigma_{\mathcal{R}}(k^2)}{\partial k^2}\bigg)\bigg\vert_{k^2=m^2}
\vert\mathcal{M}\vert^2_{\mathrm{LO}}(k^2).
\end{align}
\end{widetext}
The two terms are well expected, representing the mass correction and LSZ reduction factor known from the basic renormalization theory. 

The imaginary part $\Sigma_{\mathcal{I}}(k^2)$ enters the calculation in a more interesting way through
\begin{align}\label{eq9}
\frac{2}{\pi}\int\limits^s_0 dk^2 \vert\mathcal{M}\vert^2_{\mathrm{LO}}(k^2) \Sigma_{\mathcal{I}}(k^2)\pi^2\delta(k^2-m^2)^2.
\end{align}
Fortunately, this expression, containing the square of the delta function, is singular and ready to cancel a similar disaster elsewhere, such that only a finite part will be left. To see that explicitly, we consider the remaining cut of the diagram in Fig. \ref{fig1},
\begin{align}\label{eq10}
\includegraphics[scale=1,valign=c]{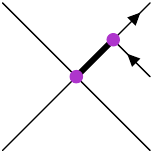}
\includegraphics[scale=1,valign=c]{cut.pdf}
\includegraphics[scale=1,valign=c]{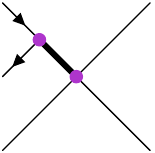}
\end{align}
leading to
\begin{align}\label{eq11}
-\frac{1}{\pi}\int\limits^s_0 dk^2 \vert\mathcal{M}\vert^2_{\mathrm{LO}}(k^2)\Sigma_{\mathcal{I}}(k^2) \bigg\vert\frac{1}{k^2-m^2+\iu\epsilon}\bigg\vert^2.
\end{align}
Adding a narrow finite width $\Gamma$ into Eq. \eqref{eq11} gives
\begin{align}\label{eq12}
\bigg\vert\frac{1}{k^2-m^2+\iu m\Gamma} \bigg\vert^2\quad\rightarrow\quad\frac{\pi}{m\Gamma}\delta(k^2-m^2).
\end{align}
With $\Sigma_{\mathcal{I}}(k^2) \rightarrow -m\Gamma$, the leading-order result of Eq. \eqref{eq2} is obtained, and the double-counting occurs. We instead suggest adding the singular expression of Eq. \eqref{eq9} to Eq. \eqref{eq11} obtaining
\begin{align}\label{eq13}
\bigg\vert\frac{1}{k^2-m^2+\iu\epsilon}\bigg\vert^2 - \frac{2\epsilon^2}{[(k^2-m^2)^2+\epsilon^2]^2}.
\end{align}
Here we used the imaginary part of Eq. \eqref{eq6} to express the $\delta$ function in Eq. \eqref{eq9}. To see the connection with Eq. \eqref{eq12}, we can multiply this formula by $\epsilon/(m\Gamma)$ and get
\begin{align}\label{eq14}
\frac{1}{m\Gamma}\bigg[\frac{\epsilon}{(k^2-m^2)^2+\epsilon^2} - \frac{2\epsilon^3}{[(k^2-m^2)^2+\epsilon^2]^2}\bigg].
\end{align}
The two terms correspond to two different representations of the same $\delta$ function, and thus, after the integration over the phase-space, they cancel each other. Therefore, contrary to common wisdom, there is no double-counting and no singularity in the end. In principle, one may ignore the contribution of Eq. \eqref{eq9} and introduce a counter-term of the same form by hand, which may be understood as an analogy of the diagram subtraction \cite{Beenakker:1996ch, Plehn:2010gp, Gavin:2013kga}. However, when going from Eq. \eqref{eq13} to Eq. \eqref{eq14}, a finite part is lost. We instead rewrite Eq. \eqref{eq13} in terms of a principal value \footnote{At this perturbative order, our result corresponds to the second term of the expansion in Eq. (2) in Ref. \cite{Tkachov:1998uy}.}
\begin{align}\label{eq15}
-\frac{\partial}{\partial k^2}\mathcal{P}\frac{1}{k^2-m^2}
\end{align}
such that the integration can be easily performed by parts, that finally gives the on-shell $\eta\bar{\eta}$ production corrections in the form of
\begin{align}\label{eq16}
\vert\mathcal{M}\vert^{2,\eta\bar{\eta}}_{\mathrm{NLO}} = 
\frac{\lambda^2_\varphi}{(4\pi)^3}\frac{\lambda^2_\eta}{2s}\bigg( \ln\bigg\vert\frac{m^2}{s-m^2}\bigg\vert -\frac{s}{m^2}\bigg).
\end{align}
The commonly used diagram subtraction scheme \cite{Beenakker:1996ch, Plehn:2010gp, Gavin:2013kga} requires subtracting
\begin{align}\label{eq17}
\vert\mathcal{M}\vert^2_{\mathrm{LO}}(m^2)\Sigma_{\mathcal{I}}(m^2) \frac{\theta(\sqrt{s}-m)}{(k^2-m^2)^2+\epsilon^2}
\end{align}
from the integrand in Eq. \eqref{eq11} with $\epsilon$ playing a role of a regulator. The so-called counter-term has been introduced to remove the resonant enhancement and prevent the reappearance of the leading-order contribution. In our simple model, for $\epsilon\rightarrow 0$ this rather ad-hoc procedure results in a similar expression as shown in Eq. \eqref{eq16}, without the second $-s/m^2$ term in the bracket -- the lost finite part mentioned earlier. Although the diagram subtraction is known to preserve the gauge invariance in a real-world theory in the limit of vanishing regulator \cite{Gavin:2013kga}, the omission of a finite part of the NLO corrections leads to a systematic error in the calculations. Using a unitarity-based fixed-order approach instead, nothing is added nor subtracted, and thus the gauge invariance should not be compromised.

\section{Cutting first! And modifying the Dyson series}

We may ask how it could happen that the contribution of Eq. \eqref{eq9}, which is crucial for the finiteness and consistency of the calculation, has been overlooked in previous studies for years. The answer can be immediately seen from Eq. \eqref{eq7} bringing the mass and LSZ corrections in a way different from how they usually appear. Usually, the Dyson series is summed first, the pole and the residuum are extracted. The summation of the Dyson series for unstable particles leads to a finite imaginary part of the propagator. Such an internal line cannot be cut \cite{Veltman:1963th}, and the singularity of Eq. \eqref{eq9} is obscured. Introducing a finite width removes the singularity of Eq. \eqref{eq11} as well. However, it has to be done consistently to prevent double-counting.

\begin{figure}
\centering\includegraphics[scale=1]{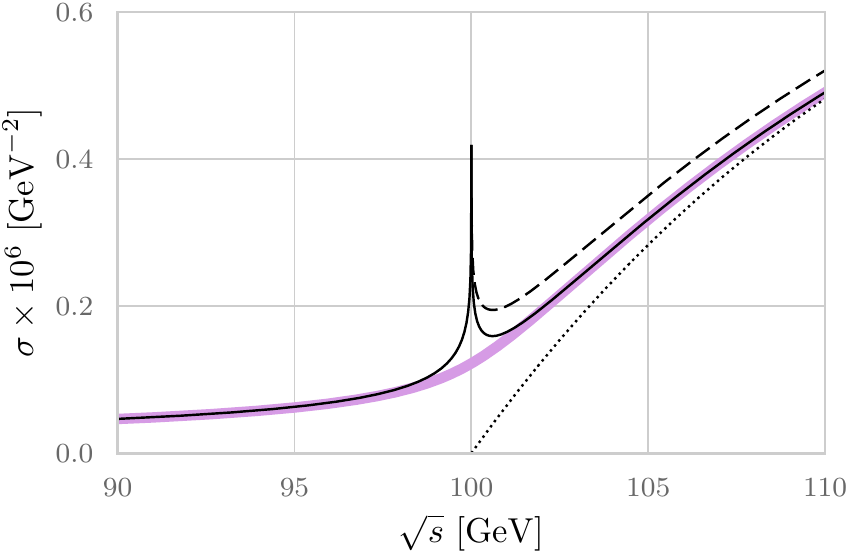}
\caption{\label{fig2} Total cross-section for  $\varphi\varphi\rightarrow\varphi\phi(\eta\bar{\eta})$ reaction with $\bar{m}=100~\mathrm{GeV}$, $\lambda_\varphi=1.3$, and $\lambda_\eta=\bar{m}\times\lambda_\varphi$, at the leading-order (dotted line), next-to-leading order in our unitarity-based fixed-order approach (solid black line), diagram subtraction (dashed line), and integrated Dyson-summed cross-section according to Eq. \eqref{eq22} (thick purple line).}
\end{figure}

Let us generalize the procedure leading to Eq. \eqref{eq16}, hopefully obtaining a description reliable even for near-threshold calculations, where higher orders have to be included systematically. Instead of summing the Dyson series for the propagator, finding its pole and residuum, we sum the series of forward diagrams similar to Fig. \ref{fig1} with any number of the $-\iu \Sigma(k^2)$ insertions in the propagator of $\phi$. Possible cuttings of such diagrams fall into two groups. First, there are two cuts analogous to those that appear in Eq. \eqref{eq3}, in which only the first or last $\phi$ line is cut. Defining
\begin{align}\label{eq18}
-\iu\tilde{\Sigma}(k^2) = \includegraphics[scale=1,valign=c]{math3a.pdf}\hskip1mm\times
\sum^\infty_{n=0}\hskip1mm\left(\includegraphics[scale=1,valign=c]{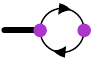}\right)^n
\end{align}
these cuts lead to the expression as in Eq. \eqref{eq5} with $\Sigma(k^2)$ replaced by $\tilde{\Sigma}(k^2)$. Second, when any of the self-energies or the connecting intermediate lines is cut, by unitarity and optical theorem, it will contribute to Eq. \eqref{eq11} by the imaginary part of $\tilde{\Sigma}(k^2)$. Summing the geometric series of Eq. \eqref{eq18}, we obtain
\begin{align}
&\tilde{\Sigma}_{\mathcal{R}}(k^2) = \frac{(k^2-m^2)^2\Sigma_{\mathcal{R}}-(k^2-m^2) (\Sigma^2_{\mathcal{R}}+\Sigma^2_{\mathcal{I}})}{(k^2-m^2-\Sigma_{\mathcal{R}})^2+ \Sigma^2_{\mathcal{I}}}\label{eq19}\\
&\tilde{\Sigma}_{\mathcal{I}}(k^2)\, = \frac{(k^2-m^2)^2\Sigma_{\mathcal{I}}}{(k^2-m^2-\Sigma_{\mathcal{R}})^2+\Sigma^2_{\mathcal{I}}}.\label{eq20}
\end{align}
And now a miracle happens! Eq. \eqref{eq19} implies
\begin{align}\label{eq21}
\tilde{\Sigma}_{\mathcal{R}}(k^2)\bigg\vert_{k^2=m^2}=0,\quad\frac{\partial\tilde{\Sigma}_{\mathcal{R}}(k^2)}{\partial k^2}\bigg\vert_{k^2=m^2} = -1.
\end{align}
Plugging this into Eq. \eqref{eq8}, the first term vanishes, while the second completely cancels the leading-order contribution of Eq. \eqref{eq2}. Therefore, \emph{the threat of double-counting is eliminated in a manifest way, and it is the leading-order part to be removed}. We note that it has been achieved without employing a specific renormalization scheme, such as the complex mass scheme, where the complex-valued counter-term removes the double-counting \cite{Denner:2014zga}. The analogue of Eq. \eqref{eq9} vanishes for $\tilde{\Sigma}_{\mathcal{I}}(k^2)$, while from Eq. \eqref{eq11} the contribution of modified Dyson series is obtained in the form of
\begin{align}\label{eq22}
-\frac{1}{\pi}\int\limits^s_0 dk^2 
\frac{\vert\mathcal{M}\vert^2_{\mathrm{LO}}(k^2)\Sigma_{\mathcal{I}}}{[k^2-m^2- \Sigma_{\mathcal{R}}(k^2)]^2+\Sigma_{\mathcal{I}}^2}
\end{align}
Including the counter-terms to absorb the ultraviolet divergence is straightforward now. We denote the renormalized mass and the real part of the self-energy as $\bar{m}$ and $\bar{\Sigma}_{\mathcal{R}}(k^2)$, respectively. Specifying the renormalization conditions as in the on-shell scheme,
\begin{align}\label{eq23}
\bar{\Sigma}_{\mathcal{R}}(\bar{m}^2) = 0,\quad \frac{\partial\bar{\Sigma}_{\mathcal{R}}(k^2)}{\partial k^2}\bigg\vert_{k^2=\bar{m}^2}=0,
\end{align}
while replacing constant $\Sigma_{\mathcal{I}}$ by $-\bar{m}\Gamma$, we obtain a usual Breit-Wigner shape integrated with the leading-order amplitude squared. 

In Fig. \ref{fig2}, the result of Eq. \eqref{eq22} is seen almost indistinguishable from the fixed-order calculation $\vert\mathcal{M}\vert^2_{\mathrm{LO}}+\vert\mathcal{M}\vert^{2,\phi}_{\mathrm{NLO}} +\vert\mathcal{M}\vert^{2,\eta\bar{\eta}}_{\mathrm{NLO}}$, except for the near-threshold region, where it smoothly interpolates what otherwise will be an unphysical peak. For some applications, such as the corrections to the thermal-averaged reaction rates in dark matter calculations, as in Ref. \cite{Harz:2012fz}, the numerical impact of this peak may be very limited. Therefore, the fixed-order calculations may be accurate enough unless we are concerned with near-threshold phenomena.

The case of nontrivial branching ratios, in our particular model, includes the effect of the $\phi\rightarrow 3\varphi$ decay and analogous contribution to the self-energy, now equal to
\begin{align}\label{eq24}
-\iu\Sigma(k^2)=\hskip1mm\includegraphics[scale=1,valign=c]{math3a.pdf}
\hskip1mm+\hskip1mm\includegraphics[scale=1,valign=c]{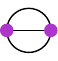}\,.
\end{align}
Summing the series of forward diagrams analogous to Fig. \ref{fig1}, with any number of the self-energy insertions of either type, leads to Eqs. \eqref{eq19} and \eqref{eq20} with $\Sigma_\mathcal{R}$ and $\Sigma_\mathcal{I}$ following from Eq. \eqref{eq24}. The imaginary part of the resulting summed $\tilde{\Sigma}(k^2)$ then receives the contributions from cuts of the $\phi$ lines connecting self-energy insertions. These cuts correspond to $\varphi\varphi\rightarrow\phi\varphi$ reaction, and have to be multiplied by the $\phi\rightarrow\eta\bar{\eta}$ branching ratio. The latter is defined as the ratio of the partial and the total decay widths -- the ratio of the imaginary parts of Eqs. \eqref{eq4} and Eq. \eqref{eq24}, respectively. Further contributions to $\tilde{\Sigma}_{\mathcal{I}}(k^2)$ come from cutting the self-energy insertions representing the $\varphi\varphi\rightarrow\eta\bar{\eta}\varphi$ and $\varphi\varphi\rightarrow4\varphi$ processes. Cutting the self-energies leads to the results proportional to the partial decay widths. Therefore, even when using Eq. \eqref{eq22} with the complete self-energy of Eq. \eqref{eq24} includes both the final states, multiplying the result by the $\phi\rightarrow\eta\bar{\eta}$ branching ratio gives the desired $\varphi\varphi\rightarrow\eta\bar{\eta}\varphi$ production rate.

Finally, it may be an instructive exercise to show that in the case of stable $\phi$ particle production (introducing $\eta$ mass larger than $m/2$), Eqs. \eqref{eq18}-\eqref{eq20} reproduce the NLO corrections, although in an unexpected way. Vanishing imaginary part of the self-energy in Eq. \eqref{eq18} implies
\begin{align}
&\tilde{\Sigma}_{\mathcal{R}}(k^2) = \Sigma_{\mathcal{R}}(k^2-m^2)\,\mathcal{P}\frac{1}{k^2-m^2-\Sigma_{\mathcal{R}}}\label{eq25}\\
&\tilde{\Sigma}_{\mathcal{I}}(k^2)\, = -\Sigma^2_{\mathcal{R}}\,\pi \delta(k^2-m^2-\Sigma_{\mathcal{R}})\label{eq26}
\end{align}
for which Eq. \eqref{eq21} remains valid, and the leading-order contribution is canceled. However, substituting Eq. \eqref{eq26} into Eq. \eqref{eq11} results into
\begin{align}\label{eq27}
\int\limits^s_0 dk^2 \vert\mathcal{M}\vert^2_{\mathrm{LO}}(k^2)\delta[k^2-m^2-\Sigma_{\mathcal{R}}(k^2)]
\end{align}
leading, up to higher-order terms, to the expressions in Eqs. \eqref{eq2} and \eqref{eq8}. We note that in this case the $\varphi\varphi\rightarrow\eta\bar{\eta}\varphi$ process has to be considered separately, as it cannot be related to the self-energy of stable $\phi$.

\section{Conclusions}

Perturbative unitarity has been used to describe resonant particle production. Employing a simple scalar model, it has been shown that the production rate is free of any singularities and double-counting, both in the fixed-order calculation and the resummed case. The former included the sum over the forward scattering diagrams with any number of self-energy insertions on the unstable particle leg followed by their cutting (extraction of the imaginary part). In the case of stable particle production, the method reproduces the well-known form of the higher-order corrections.

\begin{acknowledgements}
I would like to thank my colleagues, Tom\'a\v{s} Bla\v{z}ek and Fedor \v{S}imkovic, for their long-term support. The author was supported by the Slovak Ministry of Education Contract No. 0243/2021.
\end{acknowledgements}

\bibliographystyle{apsrev4-1.bst}
\bibliography{CLANOK.bib}

\end{document}